\documentclass[aps,prl,reprint,superscriptaddress]{revtex4-2}

\usepackage{graphicx}
\usepackage{physics}
\usepackage{xcolor}
\usepackage{float}
\usepackage[section]{placeins}
\usepackage{amsmath}

\newenvironment{psmallmatrix}
{\left(\begin{smallmatrix}}
	{\end{smallmatrix}\right)}

\begin{document}



\title{Optical signatures of noncentrosymmetric structural distortion in altermagnetic MnTe}



\normalsize

\author{Ao Wu}
\affiliation{Department of Physics, National University of Singapore, Singapore 117551, Singapore}

\author{Di Cheng}
\affiliation{Department of Physics, National University of Singapore, Singapore 117551, Singapore}

\author{Xinyun Wang}
\affiliation{Department of Physics, National University of Singapore, Singapore 117551, Singapore}

\author{Meng Zeng}
\affiliation{Department of Physics and State Key Laboratory of Quantum Functional Materials, Southern University of Science and Technology, Shenzhen, Guangdong 518055, China}

\author{Chang Liu}
\affiliation{Department of Physics and State Key Laboratory of Quantum Functional Materials, Southern University of Science and Technology, Shenzhen, Guangdong 518055, China}

\author{Xinwei Li}
\email[email: ]{xinweili@nus.edu.sg}
\affiliation{Department of Physics, National University of Singapore, Singapore 117551, Singapore}

\date{\today}
	
\begin{abstract}

The hexagonal MnTe is a prime material candidate for altermagnets, an emerging class of magnetic compounds characterized by the nontrivial interplay of antiparallel spin arrangements with their underlying crystal structures. Recognizing precise knowledge of crystal symmetry as the cornerstone of the spin-group classification scheme, we report here a native inversion-symmetry-breaking structural distortion in this compound that has previously been overlooked. Through optical polarimetry experiments and first-principle calculations, we show that MnTe belongs to the noncentrosymmetric $D_{3h}$ group, effectively resolving key inconsistencies in the earlier interpretations of Raman spectroscopy data. Our finding impacts the symmetry analysis of MnTe within the altermagnetic class and sheds light on the mechanism of its magneto-controllable N\'eel order.      


\end{abstract}

\maketitle

Understanding the symmetry properties of magnetic crystalline solids is a central element in the exploration of quantum magnetism and its emerging applications. Recent demonstrations of several unexpected transport phenomena in antiferromagnets have motivated the development of a nonrelativistic spin-group framework \cite{Naka2019, Hayami2019, Smejkal2022, Bai2022, Karube2022, Liu2022}, which introduces a new classification scheme for magnets with collinear spin orders. The most striking prediction is the identification of altermagnets \cite{ifmmodeSelseSfimejkal2022,ifmmodeSelseSfimejkal2022a,Bai2024}, a novel class of magnet crystals featuring fully compensated collinear antiferromagnetic spin arrangements that nonetheless can give rise to time-reversal symmetry-breaking phenomena typically associated with ferromagnets. 

The $\alpha$-MnTe (MnTe) is a leading material candidate for the altermagnetic class \cite{ifmmodeSelseSfimejkal2022,Mazin2023,Lovesey2023}. The compound has a hexagonal lattice structure, where Mn spins align ferromagnetically within each basal plane while the inter-plane stacking order is antiferromagnetic. The spin sublattices with opposite polarizations are positioned in a face-sharing octahedral coordination environment provided by the Te atoms, and are connected through a nonsymmorphic six-fold screw rotation, consistent with the symmetry requirements for altermagnets. Efforts have been made to explore the altermagnetic phenomenology in MnTe, with an array of notable demonstrations including the nonrelativistic spin-splitting of bands \cite{Krempasky2024,Lee2024,Osumi2024}, anomalous Hall effect (AHE) \cite{GonzalezBetancourt2023,Kluczyk2024}, chiral magnons \cite{Liu2024}, ultrafast dynamics \cite{Bossini2021,Zhu2023,Gray2024}, and time-reversal-odd N\'eel order imaging \cite{Amin2024,Hubert2025,Hariki2024}.

Despite the progress, important questions remain regarding several experimental results. Firstly, the characterization of fundamental lattice excitations, especially those revealed by Raman spectroscopy measurements, is far from complete \cite{Mobasser1985,Zhang2020,Li2022}; phonon mode assignments in obvious contradiction with the symmetry of MnTe are used for latest studies. Secondly, there is tremendous interest in the origin of the small yet finite net moment resulting from spin canting, and along with it the unexpectedly great controllability of N\'eel order by external magnetic fields \cite{Wasscher1965,Kriegner2017,Kluczyk2024}; neither phenomenon can be deduced from the lattice symmetry of MnTe, prompting the proposal of a sophisticated mechanism involving high-order spin-orbit coupling (SOC) to explain them \cite{Mazin2024}. Here, through experiment-theory comparisons, we challenge the belief that the crystal structure of MnTe is of the NiAs type [$D_{6h}$ point group, as shown in Fig.\,1(a)]. Our evidence points to a potentially delicate, yet ubiquitous noncentrosymmetric structural distortion that reduces the point group of MnTe crystals to $D_{3h}$. The newly identified crystal symmetry provides a straightforward route to resolve the above-mentioned questions.

Figure\,1(b) shows a group-theory analysis of lattice normal modes with the reported crystal structure ($D_{6h}$). Within the twelve spatial degrees of freedom contributed by two Mn atoms and two Te atoms in a unit cell, the modes should be decomposed into one $B_{1g}$ mode (silent), one $E_{2g}$ mode, two $A_{2u}$ modes, one $B_{2u}$ mode (as opposed to $B_{1u}$ found by previous analysis \cite{Mobasser1985,Onari1974}), two $E_{1u}$ modes, and one $E_{2u}$ mode. Among these, only the $E_{2g}$ mode has Raman activity. However, previous Raman spectroscopy measurements have identified more than one Raman phonon modes \cite{Zhang2020,Li2022}, which is confirmed by our measurements. 

\begin{figure}[bt]
	\centering
	\includegraphics[width=\linewidth]{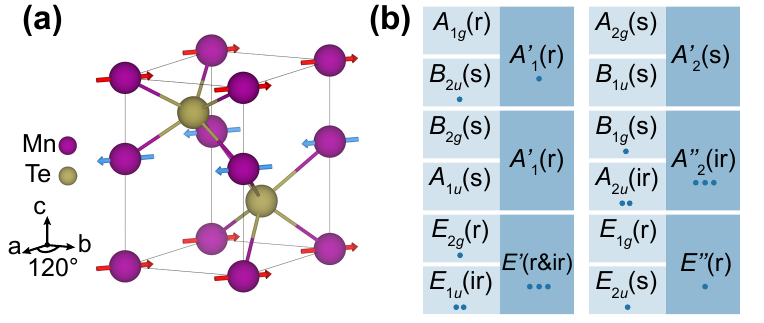}
	\caption{\small (a)~Crystal structure of MnTe. (b)~Group-theory analysis of lattice normal modes of the $D_{6h}$ structure (light blue) and the proposed $D_{3h}$ structure (dark blue). Connections between boxes show compatibility relations. There are as many modes respecting each irreducible representation as the number of dots in each box. Mode activity indicated by letters; r: Raman-active, ir: infrared-active, s: silent.}
	\label{Fig1}
\end{figure}

\begin{figure*}[tb]
	\centering
	\includegraphics[width=0.95\linewidth]{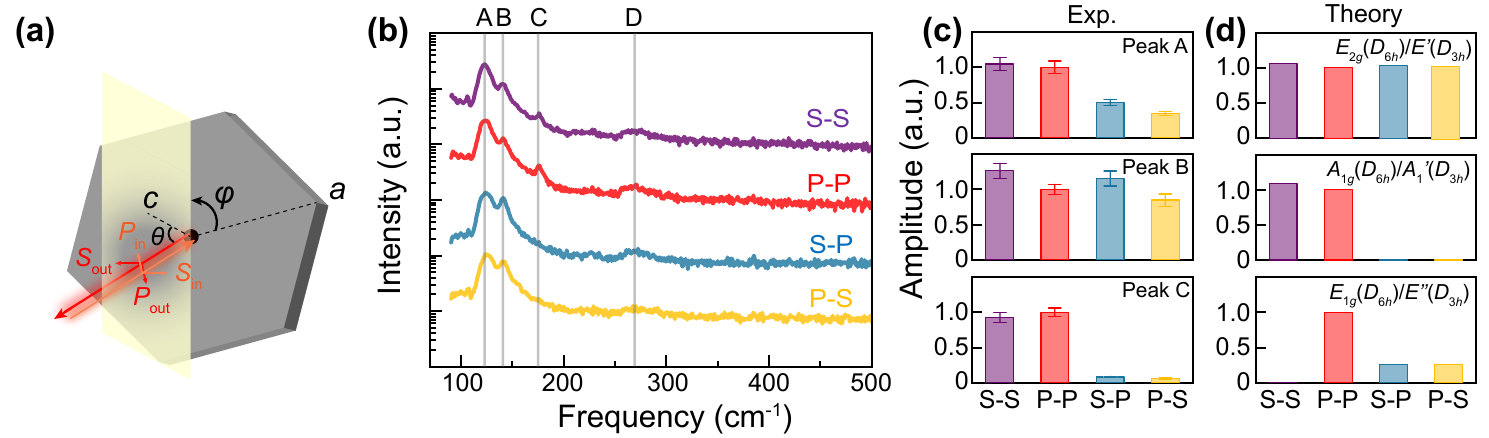}
	\caption{\small Raman spectroscopy evidence for broken inversion symmetry. (a)~Raman measurement geometry. (b)~Raman spectra for all four polarization channels at $\varphi=0$. Curves are offset for clarity. (c)~Spectral weights of peaks A-D extracted from panel (b) for all polarization channels. (d)~Theoretical prediction of scattering intensity for specified modes in the $D_{6h}$ and $D_{3h}$ groups, to be compared with, rather than fitted to, the data in panel (c). S-S, P-P, S-P, and P-S represents $S$-in $S$-out, $P$-in $P$-out, $S$-in $P$-out, and $P$-in $S$-out geometries, respectively.
	}
	\label{Fig2}
\end{figure*}

Our samples are millimeter-sized and freshly cleaved MnTe single crystals grown by the chemical vapor transport method. Powder x-ray diffraction characterizations suggest that the sample contains a minimal level of impurity in the MnTe$_2$ phase but no oxide (MnO$_2$) composition \cite{SM}. We performed  polarization-resolved Raman spectroscopy measurements at room temperature on the natural cleaving surface [0001], as shown in Fig.\,2(a). A 633~nm continuous-wave laser beam illuminated the sample at an incidence angle of $\theta=20^{\circ}$. By limiting the aperture of light passage through the microscope objective, we collected the back-scattered light tracing the same path as the incident beam but propagating in the opposite direction. The scattered light was then separated from the incident beam path through a dichroic mirror for spectral analysis. The azimuthal angle $\varphi$ separating the scattering plane (containing the crystal normal and incident beam) from the crystal $a$-axis was varied without a change in $\theta$ or a spatial shift of the beam focus.

Our Raman spectra at $\varphi=0^{\circ}$ are displayed in Fig.\,2(b), with four input-output polarization combinations: $S$-in $S$-out, $P$-in $P$-out, $S$-in $P$-out, and $P$-in $S$-out. Spectra at other $\varphi$ angles closely resemble the $\varphi=0^{\circ}$ traces \cite{SM}. Four peaks, labeled A - D, are observed. Previous studies assigned peak C (175 cm$^{-1}$) to the $E_{2g}$ phonon mode, which is the only Raman-active mode expected for the $D_{6h}$ structure, and the broad peak D (270 cm$^{-1}$) to the two-magnon continuum \cite{Mobasser1985,Zhang2020}. The origins of peak A (120 cm$^{-1}$) and peak B (140 cm$^{-1}$) remained controversial, although a possible impurity origin was proposed \cite{Zhang2020}. We deem it necessary to revise these assignments for the following reasons.

First, the observed polarization selection rule for peak C is inconsistent with an $E_{2g}$ phonon. Peak C shows pronounced spectral weights for co-polarized channels ($S$-in $S$-out and $P$-in $P$-out), but becomes extinct in cross-polarized ones ($S$-in $P$-out and $P$-in $S$-out); see Fig.\,2(b) for the spectra and Fig.\,2(c) for spectral weight analysis using Lorentzian fits to the spectra. An $E_{2g}$ phonon, however, would show similar scattering amplitudes across all polarization channels [Fig.\,2(d)], according to our calculation using the Raman intensity formula
\begin{equation}
	I_{E_{2g}} \propto |\textbf{e}_s \cdot \textbf{R}^1_{E_{2g}} \cdot \textbf{e}_i|^2+|\textbf{e}_s \cdot \textbf{R}^2_{E_{2g}} \cdot \textbf{e}_i|^2
\end{equation}
where the tabulated Raman tensors are $\textbf{R}^1_{E_{2g}}=\begin{psmallmatrix}d & 0 & 0 \\0 & -d & 0 \\0 & 0 & 0\end{psmallmatrix}$ and $\textbf{R}^2_{E_{2g}}=\begin{psmallmatrix}0 & -d & 0 \\-d & 0 & 0 \\
	0 & 0 & 0\end{psmallmatrix}$,
and $\textbf{e}_i$ and $\textbf{e}_s$ are incident and scattered polarization vectors, respectively \cite{SM}. Secondly, the assignment of peak D to two-magnon continuum originated from an early experimental work, where the peak frequency gave an estimate of the spin-spin exchange parameters \cite{Mobasser1985}. The estimate shows a large discrepancy with the parameter set derived from inelastic neutron scattering experiments \cite{Szuszkiewicz2006}, which are better probes of spin-wave dispersions. The proposed peak assignment also fails to explain the strong amplitude of peak D observed for temperatures far above the N\'eel temperature ($307$~K) of MnTe \cite{Zhang2020}, where magnons are expected to faint due to the absence of spin order. Finally, peaks A and B are unlikely to originate from impurities. The predominant impurity species in our single crystal is the MnTe$_2$ phase, but neither peak match in frequency with any Raman phonon of MnTe$_2$ \cite{Lithesis}. The fact that the amplitude ratios among peaks A-D are consistent across MnTe in dramatically different sample morphologies [bulk crystals (this work), thin films \cite{Zhang2020}, nanoflakes \cite{Li2022}] with varying levels of disorder and impurity suggest all peaks are likely to have an intrinsic origin.

\begin{figure}[b!]
	\centering
	\includegraphics[width=\linewidth]{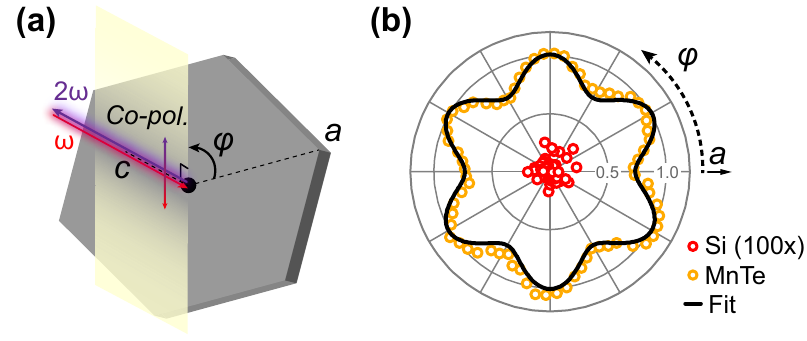}
	\caption{\small SHG evidence for broken inversion symmetry. (a) Polarization-resolved SHG measurement geometry. (b) $\varphi$-dependent SHG intensity overlaid with a fitting curve, compared with scaled-up data (by 100) from centrosymmetric silicon under identical measurement conditions.
	}
	\label{Fig3}
\end{figure}

Given that the symmetry constraints imposed by the $D_{6h}$ group are too stringent for interpreting the Raman spectra, we considered the index-two subgroups of $D_{6h}$. To resolve the correct point group, we performed second-harmonic generation (SHG) polarimetry experiments. The measurement is particularly sensitive to inversion symmetry because the second-order nonlinear susceptibility tensor $\overleftrightarrow \chi$ of the electric-dipole type, which connects nonlinear polarization $\textbf{P}^{2\omega}$ with the incident field $\textbf{E}^{\omega}$ through $P^{2\omega}_i=\chi_{ijk} E^{\omega}_jE^{\omega}_k$, must vanish for centrosymmetric groups. As shown in Fig.\,3(a), a linearly polarized pulsed laser beam (800~nm, 80~MHz, 100~fs) excited the sample at normal incidence, and the intensity of the co-polarized frequency-doubled reflected light (400~nm) was measured as a function of the azimuthal angle $\varphi$ between the polarization vector and the crystal $a$-axis.

The $\varphi$-dependent SHG intensity is plotted in Fig.\,3(b), along with the data from a (100)-cut silicon single crystal (centrosymmetric) under identical experimental conditions; their intensities differ by at least three orders of magnitude, strongly suggesting the absence of inversion symmetry in MnTe. This rules out the centrosymmetric $D_{6h}$, $C_{6h}$, and $D_{3d}$ groups. Moreover, our data contradicts with subgroups of $D_{6h}$ containing $c$-axis as a two-fold rotation axis. Two-fold rotation symmetry effectively serves as inversion symmetry within the basal plane (sign reversal of all in-plane field components), which prohibits normal-incidence SHG. This further rules out $C_{6v}$ and $D_{6}$, leaving $D_{3h}$ the only group to explain our data. Our model assuming a $D_{3h}$ structures nicely fits the petal plot \cite{SM}, from which we can determine the four nonzero elements ($\chi_{yyy}=-\chi_{xxy}=-\chi_{xyx}=-\chi_{yxx}$) within $\overleftrightarrow \chi$ contributing to the SHG intensity.



The conceived pathway of symmetry-lowering from $D_{6h}$ to $D_{3h}$ involves a distortion mode transforming as the $B_{2u}$ irreducible representation. The set of atomic displacements in the $B_{2u}$ mode features a collective vertical movement of Mn planes; see calculated results in Fig.\,4(a) and (b). Within a unit cell, a Mn layer shifts by an equal distance in the opposite direction relative to the other Mn layer carrying antiparallel spin polarizations, rendering Mn atoms no longer the centers of inversion within the MnTe$_6$ octahedra.

To examine if the microscopic crystal structure resulting from the $B_{2u}$ distortion can quantitatively describe the Raman spectra, we performed first-principle calculations of the phonon dispersions for the $D_{6h}$ and $D_{3h}$ structures [Fig.\,4(a)]. The calculations were performed using density functional perturbation theory as implemented in VASP and the Phonopy package, where we imposed the Hubbard $U$ correction and collinear antiferromagnetic order on Mn sites \cite{Mu2019,SM}. Our calculations assumed a delicate $B_{2u}$ distortion (Mn displacements $\sim$ 0.1\% of $c$-axis) and identified its impact on phonon modes.

\begin{figure*}[t!]
	\centering
	\includegraphics[width=0.9\linewidth]{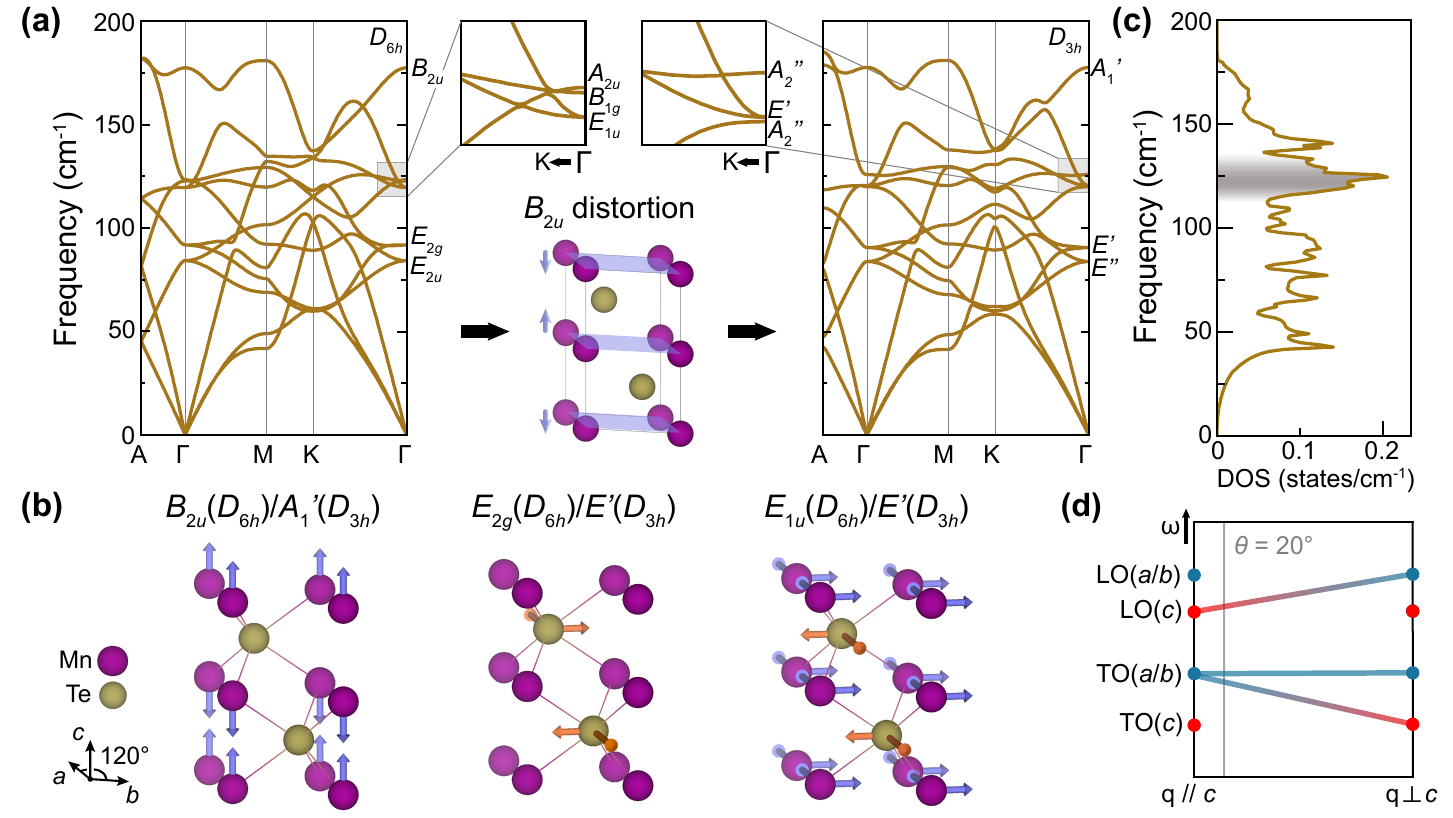}
	\caption{\small First-principle calculations of lattice normal modes. (a)~Phonon dispersion curves assuming the reported $D_{6h}$ structure (left) and the proposed $D_{3h}$ structure (right). Zone-center irreducible representations are labeled. (b)~Atomic displacements in a few labeled zone-center normal modes. (c)~Phonon density of states (DOS) spectrum. Shaded region highlights a peak structure which can give rise to two-phonon scattering. (d)~Schematic showing the evolution of TO and LO modes versus phonon wave-vector $\textbf{q}$. Our Raman measurement geometry sets the wave-vector marked by the gray line, whose intersections with colored lines give the observed modes.
	}
	\label{Fig4}
\end{figure*}

As shown in the left panel of Fig.\,4(a), the calculated frequency of the $\Gamma-$point $E_{2g}$ phonon ($D_{6h}$), which exclusively involves in-plane displacements of the heavy Te atoms [see our mode analysis in Fig.\,4(b)], is 83~cm$^{-1}$, a value too low to explain the peak C (175~cm$^{-1}$) in Raman spectra. Interestingly, upon lowering the symmetry to $D_{3h}$, the originally silent $B_{2u}$ phonon ($D_{6h}$) transforms into a Raman-active $A'_1$ phonon at 177~cm$^{-1}$ [Fig.\,4(a) right panel]. The $A'_1$ mode matches with peak C not only in frequency, but in Raman polarization selection rule as well [Fig.\,\ref{Fig2}(c) and (d)]: the Raman tensor of $A'_1$ phonons, $\textbf{R}_{A'_1}=\begin{psmallmatrix}a & 0 & 0 \\0 & a & 0 \\0 & 0 & b\end{psmallmatrix}$, lacks any off-diagonal elements and thus forbids scattering in cross-polarized channels. We therefore assign peak C to the $A'_1$ phonon.  Furthermore, having observed a pronounced maximum centered at 133~cm$^{-1}$ in the phonon density of states [Fig.\,4(c)], we assign peak D (270 cm$^{-1}$) to be a two-phonon continuum, where a pair of phonons with opposite momentum are scattered simultaneously. Its half-frequency relation with peak D along with the above-mentioned peak D's behavior above the N\'eel temperature corroborates this interpretation.

Finally, near the frequencies of the experimental peaks A and B, there are two silent $A''_2$ modes and one $E'$ mode [120~cm$^{-1}$ as shown in Fig.\,4(a) right panel]; the latter is compatible with the infrared-active $E_{1u}$ mode in $D_{6h}$, while the symmetry reduction endows the mode with additional Raman activity [Fig.\,1(b)]. The scattering intensities for the $E'$ mode, as predicted from its Raman tensors, are finite for all polarization channels \cite{SM}, aligning with the behavior of peaks A and B [Fig.\,2(c) and (d)]. Upon a close examination, we assign peaks A and B, respectively, to the transverse- and longitudinal optical (TO and LO) phonons associated with the $E'$ mode. Due to long-range dipole-dipole interactions, TO-LO frequency splitting is broadly applicable to polar phonons with Raman activity. As MnTe belongs to the uniaxial crystal class, its phonon frequency hierarchy is determined by the interplay between crystal anisotropy and electrostatic forces \cite{Loudon1964}. Referring to earlier spectroscopy results and calculations \cite{Allen1977}, we reason that the latter outweighs the former, leading to a diagram sketched in Fig.\,4(d). Given our Raman measurement configuration, which constrains the phonon propagation and polarization vectors, we identify that the two dominant modes in our spectra are: (1) TO phonon polarized within the $ab$ basal plane [TO($a$/$b$), assigned to peak A], and (2) LO phonon polarized along the $c$-axis [LO($c$), assigned to peak B]. In the Supplemental Material \cite{SM}, we present a refined calculation of the TO and LO mode polarization selection rule based on an electro-optic tensor theory, which nicely interprets the intensity variations of peaks A and B across the four polarization channels.

The reported noncentrosymmetric distortion is likely to be delicate as it has evaded detection by common structural probes such as diffraction, for which the Friedel’s law could limit the resolution power to inversion asymmetry. Nevertheless, our work has crucial implications to MnTe's altermagnetic properties. With the $D_{3h}$ structure, the six-fold screw axis no longer holds, leaving the horizontal mirror operation to connect the two magnetic sublattices. Considering this, MnTe remains altermagnetic, but would belong to a lower symmetry spin-group. In electronic band structure, this reduces the number of spin-degenerate nodal planes and activates relativistic Rashba spin-splitting. Further, as a surprising demonstration of its ultra-manipulable N\'eel-order, MnTe shows salient magnetic-field-induced hysteresis loops in AHE experiments \cite{GonzalezBetancourt2023,Kluczyk2024}. The crystal holds a small net magnetization that results from $c$-axis moment canting (estimated $\sim 10^{-5}\ \mu_\text{B}$/Mn). Despite being considered indispensible for altermagnetic spintronics applications, the origin of such canting and more importantly the sensitivity of N\'eel order to magnetic fields remains unclear. Recent theoretical work proposed a high-order SOC effect originating from a spontaneous symmetry-lowering of the structure \cite{Mazin2024}. In parallel, a staggered Dzyaloshinskii-Moriya (DM) interaction mechanism, where the DM vector switches sign between adjacent Mn layers, was also proposed \cite{Autieri2024}. With the $D_{3h}$ structure reported here, symmetry constraints on the form of DM interactions are lifted, creating new nonzero elements in the DM vector. This could turn on linear coupling channels between the net magnetization vector and the N\'eel vector \cite{Kimel2024} that would otherwise be forbidden, thereby prompting a simple yet practical model that accounts for the unusual magnetic properties.

To conclude, through group-theory analysis, optical polarimetry experiments, and first-principle calculations, we discovered a native inversion-symmetry-breaking structural distortion in $\alpha-$MnTe. The proposed structure resolves the inconsistencies in previous Raman mode assignments and points to a Dzyaloshinskii-Moriya interaction mechanism that potentially accounts for the magneto-controllability of the N\'eel-order. Our method can be generalized to a broad class of antiferromagnets with unusual symmetries for analyzing the impact of delicate structural distortions to the altermagnetic phenomenology.

\begin{acknowledgments}
We thank Jian-Rui Soh for useful discussions. X.L. and C.L acknowledge support from the SUSTech-NUS Joint Research Program (A-8002266-00-00). X.L. acknowledges support from Singapore National Research Foundation under award number NRF-NRFF16-2024-0008. Work at SUSTech was supported by the National Key R\&D Program of China (no. 2022YFA1403700).
\end{acknowledgments}

\end{document}